\documentclass[final]{ias2}

\usepackage{graphicx}
\usepackage{multirow}
\usepackage{array}

\usepackage{hyperref}

\begin{document}


\title{Nuclear Multifragmentation: Basic Concepts}

\author[sin]{G. Chaudhuri}
\email{gargi@vecc.gov.in}
\author[sin]{S. Mallik}
\email{swagato@vecc.gov.in}
\author[ain]{S. Das Gupta}
\email{dasgupta@hep.physics.mcgill.ca}
\address[sin]{Physics Group, Variable Energy Cyclotron Centre, 1/AF Bidhannagar, Kolkata 700064, India}
\address[ain]{Physics Department, McGill University, Montr{\'e}al, Canada H3A 2T8}

\begin{abstract}
We present a brief overview of nuclear multifragmentation reaction. Basic formalism of canonical thermodynamical model based on equilibrium statistical
 mechanics is described. This model is used to calculate basic observables of nuclear multifragmentation like mass distribution, fragment multiplicity, isotopic
 distribution and isoscaling. Extension of canonical thermodynamical model to a projectile fragmentation model is outlined. Application of the projectile fragmentation
model for calculating average number of intermediate mass fragments and the average size of largest cluster at different $Z_{bound}$, differential charge distribution and
 cross-section of neutron rich nuclei of different projectile fragmentation reactions at different energies are described. Application of nuclear multifragmentation reaction
in basic research as well as in other domains is outlined.
\end{abstract}

\keywords{Multifragmentation, canonical thermodynamical model, projectile fragmentation}

\pacs{25.70.Mn, 25.70.Pq}

\maketitle


\section{Introduction}
 The study of nuclear multifragmentation \cite{Dasgupta,Ono,Hartnack,Das1,Bondorf1,Gross1} is an important technique for understanding the reaction mechanism in heavy ion
 collisions at intermediate and high energies. Due to collision of projectile and target nuclei, an excited nuclear system is formed.
 If its excitation energy is greater than a few MeV/nucleon, then it breaks into many nuclear fragments of different masses. This is
 known as nuclear multifragmentation. Here 'multi' indicates 'more than two'. Generally in nuclear fission process the compound nucleus
breaks into two fission fragments. Therefore multifragmentation can be considered as the higher energy version of fission. Usually
in nuclear multifragmentation reactions, required energy of the projectile beam produced from particle accelerator varies from few MeV/nucleon to
 few GeV/nucleon. The time scales involved in nuclear multifragmentation reaction are at most of the order
 of several hundred fm/c ($1$fm/c=$3.33$x$10^{-24}$ sec).

Different theoretical models have been developed for throwing light on the nuclear multifragmentation reaction and for explaining
 the relevant experimental data. The theoretical models can be classified into two main categories: (i) Dynamical models (Boltzmann-Uehling-Uhlenbeck (BUU) model \cite{Dasgupta}, Antisymmetrised Molecular Dynamics (AMD) model \cite{Ono}, Isospin dependent quantum molecular dynamics (IQMD) model \cite{Hartnack} etc.)
 and (ii) Statistical models (Canonical Thermodynamical
 Model (CTM) \cite{Das1}, Statistical multifragmentation model (SMM) \cite{Bondorf1}, microcanonical model \cite{Gross1} etc.). In the dynamical models time evolution of the nucleons of projectile and target
 nuclei are studied whereas the statistical model calculations are based on the available phase space. Compared to dynamical models, statistical models
 are computationally much less intensive and can successfully handle different kinds of
experimental data. In this article the basic formalism of the Canonical Thermodynamical Model(CTM) and its application
for calculating mass distribution, fragment multiplicity, isotopic distribution and isoscaling are described.

Presently projectile fragmentation reaction is an important area of research  in order to study the properties of exotic nuclei. So CTM
is extended to a  model for describing the projectile fragmentation reaction. Different important observables of projectile fragmentation like intermediate mass
 fragments, largest cluster size, differential charge distribution etc are calculated from this model and compared with experimental data.

The paper is structured as follows. In section 2, we give a brief introduction of different statistical models where as the details
 of Canonical Thermodynamical Model is described in section 3 and its results are represented in section 4. The extension of CTM to
 a projectile fragmentation model is described in section 5 and some results of projectile fragmentation are explained in section 6.
 Few applications of multifragmentation are mentioned in section 7 and finally summary and conclusions are presented in section 8.

\section{Statistical Models of Multifragmentation}
Nuclear multifragmentation reactions are successfully described by statistical models based on equilibrium
 scenario of different excited fragments at freeze-out condition \cite{Das1,Bondorf1,Gross1}. In statistical models, one assumes that
 depending upon the original beam energy, the disintegrating system may undergo an initial compression and
 then begins to decompress. As the density of the system decreases,higher density regions will develop into
 composites. As this collection of nucleons begins to move outward, rearrangements, mass transfers, nuclear
 coalescence and most physics will happen until the density decreases so much that the mean free paths for
 such processes become larger than the dimension of the system.This condition is termed as freeze-out \cite{Bondorf1}.

The disintegration of excited nuclei can be studied by implementation of different statistical ensembles. Calculation by
 microcanonical ensemble is most realistic but very difficult to implement.  Usually the grand canonical
 models are easily solved and they are more commonly used. In grand canonical models total mass or total
 charge fluctuation is allowed but physically it is not allowed in intermediate energy nuclear reactions.
 Statistical multifragmentation model of Copenhagen \cite{Bondorf1}, the microcanonical models
of Gross \cite{Gross1} and Randrup and Koonin \cite{Randrup} are commonly used. Canonical Thermodynamical
 Model (CTM)\cite{Das1} was introduced later  is easier to implement analytically  and its main advantage  is that one can
 eliminate the computationally intensive Monte Carlo procedures by using the recursive technique of
 Chase and Mekzian \cite{Chase}. The results from the models based on different ensembles converge
 only under certain conditions for finite nuclei \cite{Mallik4,Mallik6}.

\section{Canonical thermodynamical model (CTM)}
Assuming that a system with $A_0$ nucleons and $Z_0$
protons at temperature $T$, has expanded to a higher than normal volume,
 the partitioning into different composites can be calculated according
to the rules of equilibrium statistical mechanics.  In a canonical model, the partitioning
is done such that all partitions have the correct $A_0, Z_0$ (equivalently
$N_0, Z_0$).

The canonical partition function is given by
\begin{eqnarray}
Q_{N_0,Z_0}=\sum\prod \frac{\omega_{I,J}^{n_{I,J}}}{n_{I,J}!}
\end{eqnarray}
Here the sum is over all possible channels of break-up (the number of such
channels is enormous); $\omega_{I,J}$
is the partition function of one composite with
neutron number $I$ and proton number $J$ respectively and $n_{I,J}$ is
the number of this composite in the given channel.
The one-body partition
function $\omega_{I,J}$ is a product of two parts: one arising from
the translational motion and another is the
intrinsic partition function of the composite:
\begin{eqnarray}
\omega_{I,J}=\frac{V}{h^3}(2\pi mT)^{3/2}A^{3/2}\times z_{I,J}(int)
\end{eqnarray}
Here
$V$ is the volume available for translational motion; $V$ will
be less than $V_f$, the volume to which the system has expanded at
break up. We use $V = V_f - V_0$ , where $V_0$ is the normal nuclear volume. For all calculations in section 4 we have considered $V_f = 6V_0$, which is obtained from experimental measurements and theoretical data fitting.

The average number of composites with $I$ neutrons and $J$ protons can be written as
\begin{eqnarray}
\langle n_{I,J}\rangle=\omega_{I,J}\frac{Q_{N_0-I,Z_0-J}}{Q_{N_0,Z_0}}
\end{eqnarray}
There are two constraints: $N_0=\sum I\times n_{I,J}$ and $Z_0=\sum J\times n_{I,J}$. Substituting eq.(3) in these two constraint conditions, two recursion relations \cite{Chase} can be obtained. Any one recursion relation can be used for calculating $Q_{N_0,Z_0}$. For example
\begin{eqnarray}
Q_{N_0,Z_0}=\frac{1}{N_0}\sum_{I,J}I\omega_{I,J}Q_{N_0-I,Z_0-J}
\end{eqnarray}
We list now the properties of the composites used in this work.  The
proton and the neutron are fundamental building blocks
thus $z_{1,0}(int)=z_{0,1}(int)=2$
where 2 takes care of the spin degeneracy.  For
deuteron, triton, $^3$He and $^4$He we use $z_{I,J}(int)=(2s_{I,J}+1)\exp(-
\beta E_{I,J}(gr))$ where $\beta=1/T, E_{I,J}(gr)$ is the ground state energy
of the composite and $(2s_{I,J}+1)$ is the experimental spin degeneracy
of the ground state.  Excited states for these very low mass
nuclei are not included.
For mass number $A=5$ and greater we use
the liquid-drop formula.  For nuclei in isolation, this reads ($A=I+J$)
\begin{eqnarray}
z_{I,J}(int)=\exp\frac{1}{T}[W_0A-\sigma(T)A^{2/3}-\kappa\frac{J^2}{A^{1/3}}
-C_s\frac{(I-J)^2}{A}+\frac{T^2A}{\epsilon_0}]
\end{eqnarray}
The expression includes the
volume energy, the temperature dependent surface energy, the Coulomb
energy and the symmetry energy.  The term $\frac{T^2A}{\epsilon_0}$
represents contribution from excited states
since the composites are at a non-zero temperature.

We also have to state which nuclei are included in computing $Q_{N_0,Z_0}$
(eq.(4)).
For $I,J$, we include a ridge along the line of stability.  The liquid-drop
formula above also gives neutron and proton drip lines and
the results shown here include all nuclei within the boundaries. The long range Coulomb interaction between
different composites is included by the Wigner-Seitz approximation\cite{Bondorf1}.

\section{Results from CTM}
Important properties of nuclear multifragmentation like mass distribution, fragment multiplicity, isotopic distribution and isoscaling are studied theoretically by using canonical thermodynamical model (CTM).

\subsection{Mass Distribution}
\begin{figure}[h]
\begin{center}
\includegraphics[width=3.5in,height=1.8in,clip]{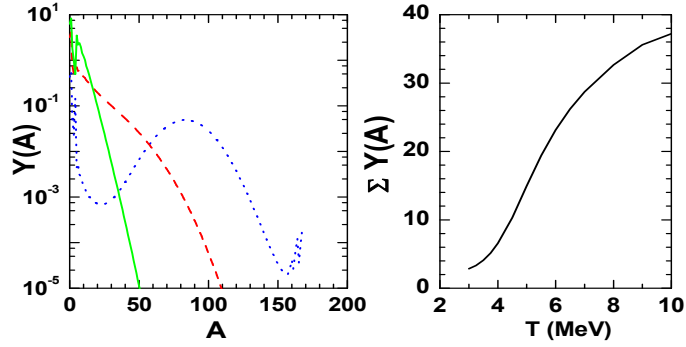}
\label{fig7}
\caption{ (Color online) Left panel: Theoretical mass distribution from $A_0=168$ and $Z_0=75$ system studied at T=3 MeV (blue dotted line), 5 MeV (red dashed line) and 7 MeV (green solid line). Right panel: Variation of total multiplicity with temperature.}
\end{center}
\end{figure}
Mass distribution of different fragments produced from the system of mass $A_0=168$ and charge $Z_0=75$ (it represents $^{112}Sn + ^{112}Sn$ central collisions after preequilibrium particle emmision), is calculated at three different temperatures and is shown in left panel of Fig. 1. At $T$=3.0 MeV (lower excitation of compound nuclear system) fission is the dominating channel i.e. the multiplicity (total number of fragments) is about 2. But at $T$=5 MeV (moderate excitation), fission channel disappears and multi-fragmentation (breaking into large number of fragments) is the dominant process with a large number of intermediate mass fragments being formed. With further increase of temperature from 5 MeV to 7 MeV (very high excitation) the system mainly breaks into a larger number of smaller mass fragments. The variation of total fragment multiplicity with temperature is shown in the right panel of Fig. 1.
\begin{figure}[h]
\begin{center}
\includegraphics[width=2.7in,height=2.7in,clip]{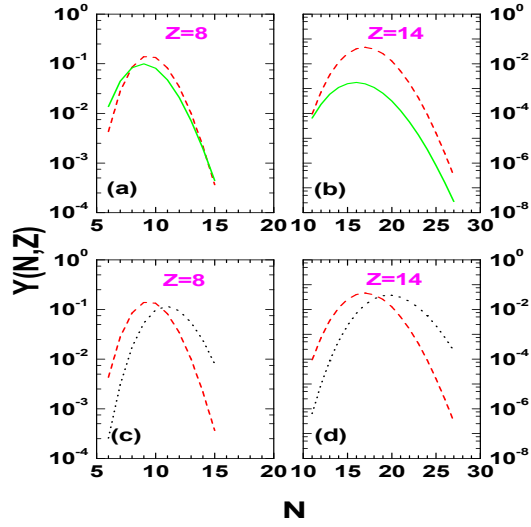}
\label{fig7}
\caption{ (Color online) Upper Panels: Theoretical isotopic distribution from $A_0=168$, $Z_0=75$ calculated at T=5 MeV (red dashed lines) and 7 MeV (green solid lines). Lower panels: Theoretical isotopic distribution from $A_0=168$, $Z_0=75$ (red dashed lines) and $A_0=186$, $Z_0=75$ (black dotted lines) calculated at T=5 MeV.}
\end{center}
\end{figure}
\subsection{Isotopic Distribution}

Isotopic distribution of $Z=8$ and $14$ fragments produced by multifragmentation of $A_0=168$ and $Z_0=75$ at two different
 temperatures $T=$ 5 and 7 MeV are shown in Fig. 2(a) and 2(b). With the increase of temperature the isotopic distributions become wider.
 Multiplicities of different isotopes having $Z=8$ and $14$ produced from two different sources of charge $Z_0=75$ and masses $A_0=168$,
$A_0=186$ is plotted in Fig. 2(c) and 2(d). From the isotopic distributions it is clear that the
production of neutron rich fragments are more from the neutron rich source $Z_0=75$, $A_0=186$ compared to the other less neutron-rich $Z_0=75$, $A_0=168$.

\subsection{Isoscaling}
Isoscaling \cite{Tsang,Chaudhuri1} is an important property for studying the symmetry energy in intermediate energy nuclear reactions .
 It is observed both theoretically and experimentally that the ratio of yields $R_{21}= Y_2(N,Z)/Y_1(N,Z)$ from two reactions $1$ and $2$
 having different isospin asymmetry ($2$ is more neutron rich than $1$) exhibit an exponential relationship as a function of neutron(N) and proton(Z) number i.e.
\begin{equation}
R_{21}= Y_2(N,Z)/Y_1(N,Z)=C\exp(\alpha N+\beta Z)
\end{equation}
where $\alpha$ and $\beta$ are isoscaling parameters and C is a normalization constant.\\
To study the isoscaling in nuclear multifragmentation, we take the dissociating systems having same $Z_1=Z_2=75$ but $A_1=168$ and $A_2=186$.
 The ratio $R_{21}$ is plotted in Fig. 3(a) as function of the neutron number for $Z=6$, $8$, $10$ and $12$ at $T=5$ MeV.  It is seen that the fragments
 produced by CTM exhibit very well the linear isoscaling behavior. The variation of the isoscaling parameter $\alpha$ with temperature in Fig. 3(b) shows that $\alpha$ gradually decreases with $T$.
$\alpha$ is related to the symmetry energy coefficient $C_s$ used in the liquid-drop formula  in eq.(5).
\begin{figure}[h]
\begin{center}
\includegraphics[width=3.6in,height=1.7in,clip]{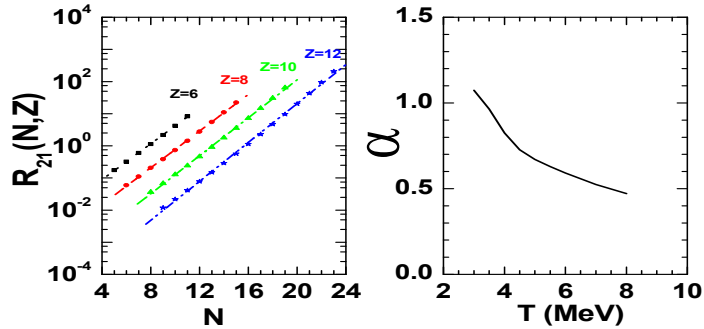}
\label{fig7}
\caption{ (Color online) Left panel:Ratios($R_{21}$) of multiplicities of the fragments $(N,Z)$ where reaction 1 is  $A_1=168$, $Z_1=75$ and reaction 2 is $A_2=186$, $Z_2=75$. Right panel: Variation of isoscaling parameter ($\alpha$) with temperature.}
\end{center}
\end{figure}

\section{Extension of CTM to a model for Projectile Fragmentation}
 Projectile fragmentation is a a very useful technique for the production of radioactive ion beam and is also important for astrophysical research.
 This led  to the extension of the canonical thermodynamical model and subsequently development into a model for projectile fragmentation \cite{Mallik2,Mallik3,Mallik101}.

The model for projectile fragmentation reaction consists of three stages: (i) abrasion, (ii) multifragmentation and (iii) evaporation. In heavy ion collision,
 if the beam energy is high enough, then in the abrasion stage at a particular impact parameter three different regions are formed: (i) projectile spectator or
 projectile like fragment (PLF) moving in the lab with roughly the velocity of the beam, (ii) participant which suffer direct violent collisions and (iii) target
 spectator or target like fragment (TLF) which have low velocities in the laboratory. Here we are interested in  the fragmentation of the PLF. Using straightline geometry average number of protons and neutrons present in the projectile spectator at different impact parameters are calculated. The total cross-section of abraded nucleus having $Z_s$ protons and $N_s$ neutrons is \cite{Mallik3,Mallik101}
\begin{equation}
\sigma_{a,N_s,Z_s}=\sum_i\sigma_{a,N_s,Z_s,T_i}
\end{equation}
where the sum is done over all impact parameter intervals and
\begin{equation}
\sigma_{a,N_s,Z_s,T_i}=2\pi \langle b_i\rangle\Delta bP_{N_s,Z_s}(\langle b_i \rangle)
\end{equation}
where $P_{N_s,Z_s}(\langle b_i \rangle)$ is the probability of formation of a projectile spectator having $Z_s$ protons and $N_s$ neutrons obtained by  using the minimal distribution
 within the impact parameter interval $\Delta b$ around $\langle b_i \rangle$ \cite{Mallik2}.

The multifragmentation stage calculation of each PLF created after abrasion at different impact parameters is done
 separately by using the Canonical Thermodynamical Model described in section 3. The impact parameter dependence of freeze-out
 temperature is considered as $T(b)=7.5-4.5(A_s(b)/A_0)$ \cite{Mallik101} where $A_s(b)$ is the mass of the projectile spectator created at impact parameter $b$ and $A_0$
is the mass number of original projectile. So freeze-out temperature of the projectile spectator is independent of the incident beam energy but it depends on the wound in the projectile.
 This parametrization of temperature profile is obtained by looking at many pieces of data from many nuclear reactions. Almost same PLF size and similar trend of temperature profile is obtained from microoscopic calculations \cite{Mallik7,Mallik9} also. The freeze-out volume in multifragmentation is $V_f(b)=3V(b)$ where $V(b)$ is the volume of projectile spectator created at $b$. Using CTM for an abraded system $N_s,Z_s$ at temperature $T_i$ average population of the composite with neutron number $n$, proton number $z$ is calculated in the multifragmentation stage. Denoting this by $M_{n,z}^{N_s,Z_s,T_i}$ and summing over all the abraded $N_s,Z_s$ that can yield $n,z$, the primary cross-section for $n,z$ is
\begin{equation}\\
\sigma_{n,z}^{pr}=\sum_{N_s,Z_s,T_i}M_{n,z}^{N_s,Z_s,T_i}
\sigma_{a,N_s,Z_s,T_i}
\end{equation}

The excited fragments produced after multifragmentation decay to their stable ground states. Its can $\gamma$-decay to shed its energy but may also decay by light
particle emission to lower mass nuclei.  We include emissions of $n,p,d,t,^3$He and $^4$He. Particle decay widths are obtained using the Weisskopf's evaporation theory
 \cite{Weisskopf}. Fission is also included as a de-excitation channel though for the nuclei of mass $<$ 100 its role will be quite insignificant. Details of the
 implementation of evaporation model can be found in \cite{Mallik1}

\section{Results from Projectile Fragmentation Reactions}
The projectile fragmentation model is used to calculate the basic observables of projectile fragmentation like the average number of intermediate mass fragments
 ($M_{IMF}$), the average size of the largest cluster and their variation with bound charge ($Z_{bound}$), differential charge distribution,cross-section of neutron
 rich fragments for different nuclear reactions at intermediate energies with different projectile target combinations.

\subsection{$M_{IMF}$ variation with $Z_{bound}$}
\begin{figure}[h]
\begin{center}
\includegraphics[height=2.0in,width=3.6in]{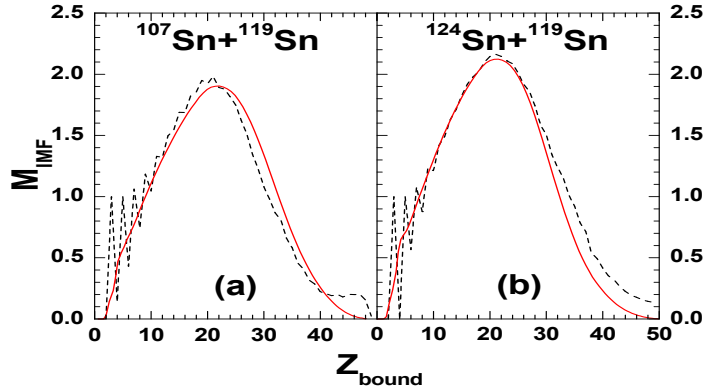}
\label{fig3}
\caption{ (Color online) Mean multiplicity of intermediate-mass fragments $M_{IMF}$, as a function of $Z_{bound}$ for (a) $^{107}$Sn on $^{119}$Sn and
(b) $^{124}$Sn on $^{119}$Sn reaction obtained from projectile fragmentation model (red solid lines). The experimental results are shown by the black dashed lines. }
\end{center}
\end{figure}
The variation of the average number of intermediate mass fragments $M_{IMF}$ ($3{\le}Z{\le}20$) with $Z_{bound}$ (=$Z_s$ minus charges of all composites
with charge $Z=1$) for $^{107}$Sn on $^{119}$Sn and $^{124}$Sn on $^{119}$Sn reactions is shown in Fig.4. The theoretical calculation reproduces the average
 trend of the experimental data very well. The experiments are done by ALADIN collaboration in GSI at 600A MeV \cite{Ogul}. At small impact parameters,
the size of the projectile spectator (also $Z_{bound}$) is small and the temperature of the dissociating system is very high. Therefore the PLF will break
 into fragments of small charges (mainly $Z=1, 2$). Therefore the IMF production is less. But at mid-central collisions PLF's are larger in size and the
temperature is smaller compared  to the  previous case, therefore larger number of IMF's are produced. With further increase of impact parameter, though
the PLF size (also $Z_{bound}$) increases,  the temperature is low, hence breaking of dissociating system is very less (large fragment remains) and therefore IMF production is less.

\subsection{Differential charge distribution}
\begin{figure}[h]
\begin{center}
\includegraphics[height=2.0in,width=3.6in]{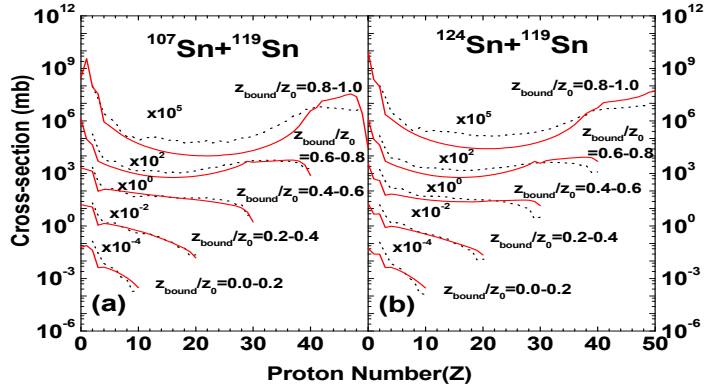}
\label{fig4}
\caption{ (Color online) Theoretical differential charge cross-section distribution (red solid lines) for (a) $^{107}$Sn on $^{119}$Sn and (b)
 $^{124}$Sn on $^{119}$Sn reaction compared with the experimental data (black dashed lines).}
\end{center}
\end{figure}
The differential charge distributions for different intervals of $Z_{bound}/Z_0$ are calculated by the projectile fragmentation model for $^{119}$Sn
and $^{124}$Sn on $^{119}$Sn reactions and compared with experimental data \cite{Ogul}. This is shown in Fig.5. For the sake of clarity the distributions
 are normalized with different multiplicative factors. At peripheral collisions (i.e. $0.8{\le}Z_{bound}/Z_0{\le}1.0$)  due to small temperature of PLF,
it breaks into one large fragment and small number of light fragments, hence the charge distribution shows $U$ type nature. But with the decrease of impact
parameter the temperature increases, the PLF breaks into larger number of fragments and the charge distributions become steeper. The features of the data are nicely
reproduced by the model.

\subsection{Size of largest cluster and its variation with $Z_{bound}$}
\begin{figure}[h]
\begin{center}
\includegraphics[height=2.0in,width=3.6in]{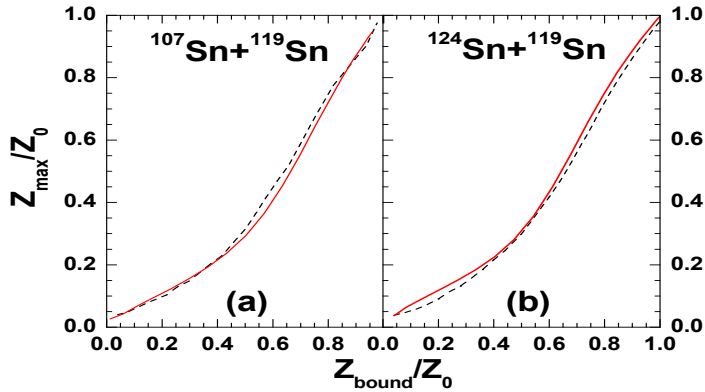}
\label{fig5}
\caption{ (Color online) $Z_{max}/Z_0$ as a function of $Z_{bound}/Z_0$ for (a) $^{107}$Sn on $^{119}$Sn and (b) $^{124}$Sn on $^{119}$Sn reaction
obtained from projectile fragmentation model (red solid lines). The experimental results are shown by the black dashed lines. }
\end{center}
\end{figure}
Average size of the largest cluster produced at different $Z_{bound}$ values is calculated in the framework of projectile fragmentation model
 for $^{119}$Sn and $^{124}$Sn on $^{119}$Sn reactions.  In Fig.6 the variation of $Z_{max}/Z_0$ ($Z_{max}$ is the average number of proton
 content in the largest cluster) with $Z_{bound}/Z_0$ obtained from theoretical calculation and experimental result are shown. Very nice agreement
 with experimental data is observed.

\subsection{Cross-section and binding energy of neutron rich nuclei}
Projectile fragmentation cross-sections of many neutron-rich isotopes have been measured experimentally from the $^{48}Ca$ and $^{64}Ni$ beams at
 $140$ MeV per nucleon on $^{9}Be$ and $^{181}Ta$ targets \cite{Mocko}. Our theoretical model reproduces the cross-sections of projectile fragmentation
 experiments very well \cite{Mallik2,Mallik3,Mallik101}. A remarkable feature is the co-relation between the measured fragment cross-section ($\sigma$)
 and the binding energy per nucleon($B/A$). This observation has prompted attempts of parametrization of cross-sections \cite{Mocko2,Mocko3,Chaudhuri2}.
One very successful parametrization is
\begin{equation}\\
\sigma=Cexp[\frac{B}{A}\frac{1}{\tau}]
\end{equation}
\begin{figure}[h]
\begin{center}
\includegraphics[width=2.7in,height=2.1in,clip]{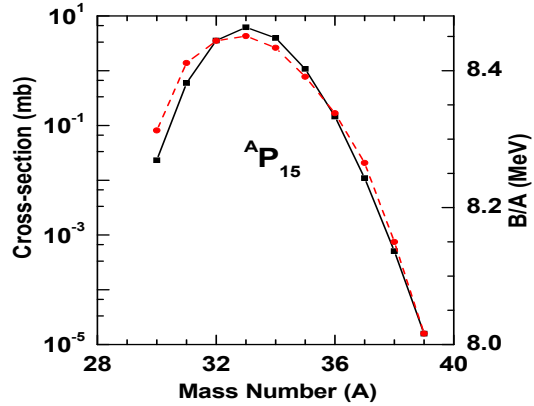}
\label{fig7}
\caption{ Fragment cross-section (circles joined by red dotted line) for  $^{64}Ni$ on $^{9}Be$ reaction and binding energy per nucleon
 (squares joined by black solid line) plotted as a mass number for $Z=15$ isotopes.}
\end{center}
\end{figure}
Here $\tau$ is a fitting parameter. In this parametrization we have not considered the pairing energy contribution in nuclear binding energy.
Here we have calculated production cross-sections $Z=15$ isotopes for $^{64}Ni$ on $^{9}Be$ reaction from projectile fragmentation model and plotted
 in log scale in Fig.-7 (circles joined by red dotted line). The variation of the theoretical binding energy per nucleon for same isotopes of $Z=15$
in linear scale is also shown in the same figure (squares joined by black solid line). The similar trend of the cross-section curve (in log scale) and
binding energy curve (in linear scale) confirms the validity of above parametrization from our model. By this method we can interpolate (or extrapolate)
 the cross-section of an isotope if the binding energy is known.  We can also estimate the binding energy of an isotope by measuring its cross-section experimentally..

\section{Application of Nuclear Multifragmentation}
Nuclear multifragmentation is very useful for studying nuclear liquid gas phase transition and for investigation of nuclear matter
 at sub-saturation densities. Projectile fragmentation is very useful technique for production of radioactive ion beam and is useful for nuclear structure studies as well as
 for astrophysical research. Nuclear multifragmentation can be used for spallation reaction (nuclear power production), nuclear waste management (environment protection),
 proton and ion therapy (medical applications), radiation protection of space missions (space research) etc. Thus nuclear multifragmentation is an important
tool in basic research as well as in a wide variety of other applications.

\section{Summary}
The study of nuclear multifragmentation is an important area of research in intermediate energy heavy-ion collisions. The canonical
thermodynamical model which is based on analytic evaluation of the partition function has been used to calculate different observables characterizing
the multifragmentation reaction and some simple results are displayed. This simple analytical model is extended to develop a model for projectile fragmentation
 which is important for the study of exotic nuclei as well as for astrophyical research. A few typical observables like average multiplicity of intermediate mass fragments, differential charge distribution, cross-section of neutron rich nuclei, size of the largest cluster and its variation with $Z_{bound}$ are calculated using this model and their good agreement with experimental data
confirms the justification of the assumptions made in the model. Apart from basic research, the disintegration of the excited nuclei into many pieces also finds its application in a wide variety of other fields.

\bibliographystyle{pramana}
\bibliography{references}

\end{document}